\documentclass[prl,aps,preprintnumbers,twocolumn]{revtex4}

\usepackage{graphicx}
\usepackage{epsfig}
\begin{document}
\preprint{ANL-HEP-PR-07-52}

\title{Angular distribution of leptons from the decay of massive
  vector bosons}

\author{Edmond L. Berger$^{1}$, Jian-Wei Qiu$^{1,2}$, 
       and Ricardo A. Rodriguez-Pedraza$^2$}
\affiliation{
${}^1$High Energy Physics Division, 
      Argonne National Laboratory, 
      Argonne, IL 60439, U.S.A.\\
${}^2$Department of Physics and Astronomy, 
      Iowa State University,
      Ames, IA 50011, U.S.A.
}

\begin{abstract}
We examine the transverse momentum $Q_\perp$ 
dependence of the helicity structure 
functions for massive vector bosons of mass $Q$ in hadron reactions.  
We demonstrate that large logarithmic terms of the form 
$\ln(Q/Q_\perp)$ in the helicity structure 
functions have the same origin as the logarithmic terms in the 
angular-integrated cross section and that they can be resummed to 
all orders in the strong coupling $\alpha_s$, in the same way as the 
angular-integrated cross section.  
As a consequence of current conservation, 
the resummed part of the helicity structure functions preserves the 
Lam-Tung relation as a function of $Q_\perp$ 
to all orders in $\alpha_s$. 
\end{abstract}
\date{August 30, 2007}

\maketitle

{\bf Introduction\ }
Massive vector boson production in hadronic interactions is a potent
probe of short-distance dynamics in strong and electroweak
interactions.  The production of massive virtual photons, the
Drell-Yan process~\cite{dy}, tests predictions of perturbative quantum
chromodynamics (QCD) at high orders in perturbation theory and
constrains parton distributions.  Production of the intermediate weak
vector bosons, the $W$ and $Z$, supplies accurate measurements of the
masses of these bosons, and, particularly in the case of the $W$
boson, provides electroweak observables that bound the mass of the
Higgs boson within the framework of the standard model of particle
physics \cite{tevatron-w-higgs}. 

The massive virtual photon, the $W$ boson, and the $Z$ boson all have
important decay modes into pairs of leptons.  The angular distribution
of these leptons, measured in the rest-frame of the parent states,
determines the alignment (polarization) of the massive vector boson
and, consequently, more precise information on the production dynamics
than is accessible from the spin-averaged rate alone.  An understanding 
of the changes expected in the angular distribution as a function 
of the transverse momentum $Q_{\perp}$ 
is a topic of considerable importance, both for refined 
tests of QCD and for reduced systematic uncertainties on the determination 
of the W boson mass \cite{bqy-wz,Ellis:1997sc}.  
In this Letter
, we derive predictions of the angular distribution as a
function of $Q_{\perp}$, including resummation to all orders  
in the strong coupling strength $\alpha_s$ of the effects of initial
state soft gluon radiation.

The general formalism for the description of the angular distribution in 
terms of helicity structure functions is developed for the Drell-Yan process 
in Ref.~\cite{lt-dy}.  The cross section may be expressed as 
\begin{eqnarray}
\frac{d\sigma}{d^4q d\Omega}
&=&
\frac{\alpha_{\rm em}^2}{2(2\pi)^4 S^2 Q^2}
\left[W_T (1+\cos^2\theta) \right.
\nonumber\\
&& 
+ W_L (1-\cos^2\theta) + W_{\Delta} (\sin(2\theta)\cos\phi) 
\nonumber\\
&& 
\left.
+ W_{\Delta\Delta} (\sin^2\theta\cos(2\phi)) \right]\, , 
\label{x-sec-angular}
\end{eqnarray}
where $q$ denotes the four-vector momentum of the massive virtual photon. The 
variables $\theta$ and $\phi$ are the polar and azimuthal angles
in the virtual photon rest frame, and $\Omega$ is the solid angle
$\Omega(\theta,\phi)$.  The four independent helicity structure
functions $W_i$ depend on the virtual photon's mass $Q$, transverse
momentum $Q_\perp$, and rapidity $y$, as well as on the center-of-mass
energy $\sqrt{S}$ of the production process.  The dependence of the 
four $W_i$ on $Q_\perp$ is our central focus in this Letter.   

At the most basic level, a massive virtual photon arises from
quark-antiquark annihilation $q+\bar{q}\to \gamma^*$ in a collision
of hadrons, and it is produced with $Q_\perp=0$ within
the framework of collinear factorization in perturbative
QCD~\cite{css-fac}.  First-order gluon radiation supplies finite $Q_\perp$, 
through the quark-antiquark and quark-gluon
subprocesses, $q+\bar{q}\to\gamma^*+g$ and $q+g\to\gamma^*+q$.  These
finite-order subprocesses yield {\it singular} differential cross
sections as a function of $Q_\perp$ in the limit $Q_\perp\to
0$. For the angular-integrated cross section, $d\sigma/d^{4}q$,
it is well established that this divergence can be removed after resummation 
of the singular terms from initial-state gluon emission to all orders 
in $\alpha_s$~\cite{DDT-qt,css-resum}.  
In this Letter, we examine in detail the structure of the 
singular contributions for each helicity structure function $W_i$.  We 
show explicitly  
that the $Q_\perp \rightarrow 0$ singular contributions to $W_T$, $W_L$, and 
$W_{\Delta\Delta}$ have the same origin as that of the angular
integrated cross section and that they can be resummed to all orders 
in $\alpha_s$.  

{\bf Asymptotic tensor\ } 
In this section, we extract the perturbatively 
singular expressions for each term in the Drell-Yan hadronic tensor, and we 
introduce an asymptotic tensor that is defined to include all the singular 
terms and, in addition, to conserve the electromagnetic current. 

The full Drell-Yan hadronic tensor can be written as
\cite{lt-dy}
\begin{eqnarray}
W^{\mu\nu} 
&=& 
- \left(g^{\mu\nu}-T^\mu T^\nu\right) 
  \left(W_T + W_{\Delta\Delta}\right)
\nonumber \\
&&
- 2 X^\mu X^\nu W_{\Delta\Delta} 
+ Z^\mu Z^\nu \left( W_L - W_T - W_{\Delta\Delta} \right)
\nonumber \\
&&
- \left( X^\mu Z^\nu + X^\nu Z^\mu \right) W_{\Delta}\, , 
\label{W-q0sfs}
\end{eqnarray}
where $T^\mu=q^\mu/Q$, and $X^\mu$, $Y^\mu$, and $Z^\mu$ are orthogonal
unit vectors in the virtual photon's rest frame, with $T^2=1$, and 
$X^2=Y^2=Z^2=-1$.  Different choices 
of the axes lead to different $\vec{q}=0$ frames~\cite{lt-dy}. 
In the Collins-Soper frame \cite{cs-frame}, all perturbatively
calculated helicity structure functions at order $\alpha_s$ and 
beyond are singular as $Q_\perp/Q\to 0$: 
$W_T$ and $W_\Delta$ have the power divergences, 
$Q^2/Q_\perp^2$ and $Q/Q_\perp$, respectively, 
as well as $\ln(Q/Q_\perp)$ divergences,
while $W_L$ and $W_{\Delta\Delta}$ show only
$\ln(Q/Q_\perp)$ divergences 
\cite{Collins:1978yt,bv-dy}.
By expanding the full Drell-Yan hadronic tensor in Eq.~(\ref{W-q0sfs}) 
in the limit $Q_\perp/Q\to 0$, we obtain the following form for its 
singular terms \cite{rr-thesis,bqr-long}
\begin{eqnarray}
W^{\mu\nu}_{\rm Sing}
&=& 
\left(-g^{\mu\nu}+\bar{n}^\mu n^\nu + n^\mu \bar{n}^\nu
\right) W_2^{\rm Asym}
\nonumber \\
&& + 
\frac{1}{\sqrt{2}} 
\left[ \frac{Q_\perp}{Q} 
       \left(n_\perp^\mu \bar{n}^\nu +
             \bar{n}^\mu n_\perp^\nu\right)\, {\rm e}^y 
\right] 
\nonumber \\
&& \hskip 0.2in \times
\left(W_2^{\rm Asym} - \frac{Q}{Q_\perp} W_1^{\rm Asym} \right)
\nonumber\\
&& + 
\frac{1}{\sqrt{2}} 
\left[ \frac{Q_\perp}{Q} 
       \left(n_\perp^\mu n^\nu +
             n^\mu n_\perp^\nu\right)\, {\rm e}^{-y} 
\right] 
\nonumber \\
&& \hskip 0.2in \times
\left(W_2^{\rm Asym} + \frac{Q}{Q_\perp} W_1^{\rm Asym} \right)
\label{w-singular}
\end{eqnarray}
with two asymptotically divergent scalar functions:
$W_2^{\rm Asym} \propto Q^2/Q_\perp^2$ and 
$W_1^{\rm Asym} \propto Q/Q_\perp$ as 
$Q_\perp/Q\to 0$.  
In Eq.~(\ref{w-singular}), $\bar{n}^\mu = \delta^{\mu +}$,
$n^\mu = \delta^{\mu -}$, and $n_\perp^\mu = \delta^{\mu\perp}$
are unit vectors that specify the light-cone coordinates of the 
laboratory frame, with $n^2 = \bar{n}^2 = 0$, 
$n^2_\perp = -1$, $n\cdot\bar{n} = 1$, and 
$n_\perp\cdot n = n_\perp\cdot\bar{n} = 0$.
In this frame, the incoming hadron momenta are 
$P_1^\mu = \sqrt{S/2}\,\bar{n}^\mu$ and 
$P_2^\mu = \sqrt{S/2}\, n^\mu$, respectively, and the virtual
photon momentum:
$q^\mu=Q^+\,\bar{n}^\mu + Q^-\,n^\mu + Q_\perp\,n_\perp^\mu$, with 
$Q^+ = \sqrt{(Q^2+Q_\perp^2)/2}\, e^{y}$ and 
$Q^- = \sqrt{(Q^2+Q_\perp^2)/2}\, e^{-y}$.

The singular tensor as $Q_\perp/Q\to 0$ in 
Eq.~(\ref{w-singular}) is not current conserving since 
$q_\mu\, W^{\mu\nu}_{\rm Sing} \neq 0$. In order to 
resum the singular terms of the hadronic tensor to all orders
in $\alpha_s$, we require a tensor that includes
all the singular terms {\em and} also conserves the current
perturbatively at any order of $\alpha_s$.  We use the term  
{\em asymptotic tensor} for this current-conserving tensor.  
We define it to be 
\begin{eqnarray}
W^{\mu\nu}_{\rm Asym}
&=& 
\left(-g^{\mu\nu}+\bar{n}^\mu n^\nu + n^\mu \bar{n}^\nu
\right) W_2^{\rm Asym}
\nonumber \\
&& + 
\frac{Q_\perp}{Q^-} 
\left(n_\perp^\mu \bar{n}^\nu + \bar{n}^\mu n_\perp^\nu
     +\frac{Q_\perp}{Q^-} \bar{n}^\mu \bar{n}^\nu \right) 
\nonumber \\
&& \hskip 0.2in \times
\frac{1}{2}
\left[W_2^{\rm Asym} - \frac{Q}{Q_\perp} W_1^{\rm Asym} \right]
\nonumber\\
&& + 
\frac{Q_\perp}{Q^+} 
\left(n_\perp^\mu n^\nu + n^\mu n_\perp^\nu
     +\frac{Q_\perp}{Q^+} n^\mu n^\nu \right) 
\nonumber \\
&& \hskip 0.2in \times
\frac{1}{2}
\left[W_2^{\rm Asym} + \frac{Q}{Q_\perp} W_1^{\rm Asym} \right]\, .
\label{w-asym}
\end{eqnarray}
It is equal to the singular tensor in Eq.~(\ref{w-singular}) 
plus a minimal non-singular term such that  
$q_\mu\, W^{\mu\nu}_{\rm Asym} = 0$.

The angular-integrated Drell-Yan cross section is obtained from the trace,
$d\sigma/d^4q \propto -g_{\mu\nu}\,W^{\mu\nu}$.   The trace of the
asymptotic tensor in Eq.~(\ref{w-asym}) should therefore be fixed by 
the asymptotic term $W^{\rm Asym}$ in the angular-integrated
Drell-Yan transverse momentum distribution~\cite{css-resum}. This 
statement allows us to fix uniquely the asymptotically 
divergent function $W_2^{\rm Asym}$ in Eq.~(\ref{w-asym}):  
$W_2^{\rm Asym} = W^{\rm Asym}/2$.  However, the angular-integrated Drell-Yan 
cross section cannot fix the second scalar function $W_1^{\rm Asym}$ in 
Eq.~(\ref{w-asym}). This second function represents the singular perturbative 
behavior of the structure function $W_\Delta$.  We can set this question 
aside for the purposes 
of the present Letter and concentrate on the resummation of the singular 
terms of the other three helicity structure functions, $W_T$, $W_L$, and 
$W_{\Delta\Delta}$.

We reexpress the asymptotic tensor in terms of the previously 
defined unit vectors in the Collins-Soper frame as 
\begin{eqnarray}
W^{\mu\nu}_{\rm Asym}
&=& 
\left[
\left(-g^{\mu\nu}+T^\mu T^\nu\right)
-\frac{Q_\perp^2/Q^2}{1+Q_\perp^2/Q^2}\, X^\mu X^\nu
\right. \nonumber \\
&& \hskip 0.2in \left.
-\frac{1}{1+Q_\perp^2/Q^2}\, Z^\mu Z^\nu
\right]
\frac{W^{\rm Asym}}{2}
\nonumber \\
& - & 
\frac{1}{1+Q_\perp^2/Q^2}\,
\left[ X^\mu Z^\nu + Z^\mu X^\nu \right]
W_1^{\rm Asym}\, .
\label{w-asym-cs}
\end{eqnarray}
Upon comparison with Eq.(2), we immediately derive the 
corresponding asymptotic 
helicity structure functions,
\begin{eqnarray}
W_T^{\rm Asym}
&=& 
\left(1-\frac{1}{2} \frac{Q_\perp^2/Q^2}{1+Q_\perp^2/Q^2} \right)
\frac{W^{\rm Asym}}{2}
\approx
\frac{W^{\rm Asym}}{2}\, ,
\nonumber \\
W_L^{\rm Asym}
&=& 
\frac{Q_\perp^2/Q^2}{1+Q_\perp^2/Q^2}\,
\frac{W^{\rm Asym}}{2}
\approx
\frac{Q_\perp^2}{Q^2}\,
\frac{W^{\rm Asym}}{2}\, ,
\label{st-asym-cs} \\
W_{\Delta\Delta}^{\rm Asym}
&=& 
\frac{1}{2}\,\frac{Q_\perp^2/Q^2}{1+Q_\perp^2/Q^2}\,
\frac{W^{\rm Asym}}{2}
\approx
\frac{1}{2}\,\frac{Q_\perp^2}{Q^2}\,
\frac{W^{\rm Asym}}{2}\, .
\nonumber
\end{eqnarray}
Our first result is this derivation that the asymptotic part of 
these three helicity structure functions is proportional directly 
to the asymptotic part of the angular-integrated 
Drell-Yan transverse momentum distribution.

{\bf Perturbatively finite tensor\ } 
We show in this section that the three asymptotic  
helicity structure functions derived in last section include 
all the divergent terms of the corresponding 
perturbatively calculated helicity structure functions, and 
therefore, that we can define a perturbatively finite tensor 
\begin{equation}
W^{\mu\nu}_{\rm Finite} \equiv
W^{\mu\nu}_{\rm Pert} - W^{\mu\nu}_{\rm Asym}\, ,
\label{w-finite}
\end{equation}
at any order of $\alpha_s$.  This perturbative finite tensor
conserves the current since the asymptotic tensor conserves
the current.

In terms of QCD collinear factorization \cite{css-fac}, we 
can express the helicity structure functions in terms of 
parton-level helicity structure functions, $w_i$:
\begin{equation}
W_i
= \sum_{ab}
\int \frac{d\xi_1}{\xi_1}  \int \frac{d\xi_2}{\xi_2}\,
\phi_a(\xi_1)\, \phi_b(\xi_2)\,
w_i(\xi_1,\xi_2,q)
\label{W-fac}
\end{equation}
with $i=T,L,\Delta\Delta$, and incoming parton distributions
$\phi_f(\xi)$ of flavor $f$ and momentum fraction $\xi$.  We
have the same factorization relation for 
$W^{\rm Asym}$~\cite{css-resum}. 

For the spin-averaged $2\to 2$ scattering processes 
$q\bar{q} \rightarrow \gamma^* g$ and $qg \rightarrow \gamma^* q$,  
the asymptotic terms at the parton-level are~\cite{bqr-long},
\begin{eqnarray}
\frac{w_{q\bar{q}}^{\rm Asym}}{2}
&\approx & 
e_q^2\,\frac{8\pi^2\alpha_s}{3x_1x_2}\,\frac{Q^2}{Q_\perp^2}
\Bigg\{
P_{qq}(z_2)\delta(1-z_1)
\nonumber \\
&& \hskip 0.9in
+P_{qq}(z_1)\delta(1-z_2)
\nonumber \\
&& 
+2\, C_F\delta(1-z_1)\delta(1-z_2)
\left[\ln(\frac{Q^2}{Q_\perp^2})-\frac{3}{2}\right]
\Bigg\}\, ;
\nonumber\\
\frac{w_{qg}^{\rm Asym}}{2} 
&\approx & 
e_q^2\,\frac{8\pi^2\alpha_s}{3x_1x_2}\,\frac{Q^2}{Q_\perp^2}\,
P_{qg}(z_2)\delta(1-z_1)\, ,
\label{w-asym-parton}
\end{eqnarray}
with $x_1=Q/\sqrt{S}\,e^y$ and $x_2=Q/\sqrt{S}\,e^{-y}$.
The parton-to-parton splitting functions are 
\begin{eqnarray}
P_{qq}(z) 
&=& C_F 
\left[ 
\frac{1+z^2}{(1-z)_+} + \frac{3}{2}\delta(1-z)
\right]\, ,
\label{pqq}
\\
P_{qg}(z) 
&=& T_R 
\left[ z^2 + (1-z)^2 \right]\, ,
\label{pqg}
\end{eqnarray}
with $C_F=(N_c^2-1)/(2N_c)=4/3$, $T_R=1/2$.
Using Eq.~(\ref{st-asym-cs}) we find that as $Q_\perp/Q\to 0$, 
the partonic-level asymptotic terms in Eq.~(\ref{w-asym-parton}) 
remove all divergent contributions of the corresponding 
perturbatively calculated helicity structure functions. For 
the quark-antiquark annihilation subprocess,  
\begin{eqnarray}
w^{q\bar{q}}_T 
&-& \frac{w_{q\bar{q}}^{\rm Asym}}{2}\
\Rightarrow \,
{\cal O}(Q_\perp^0)
\nonumber \\
w^{q\bar{q}}_L 
&-& \frac{Q_\perp^2}{Q^2}\,
  \frac{w_{q\bar{q}}^{\rm Asym}}{2}\
\Rightarrow \,
{\cal O}(Q_\perp^2)
\nonumber \\
w^{q\bar{q}}_{\Delta\Delta} 
&-& \frac{1}{2}\,\frac{Q_\perp^2}{Q^2}\,
  \frac{w_{q\bar{q}}^{\rm Asym}}{2}\
\Rightarrow \,
{\cal O}(Q_\perp^2) . 
\label{sf-finite-qq-qt0}
\end{eqnarray}
For the quark-gluon subprocess,  
\begin{eqnarray}
w^{qg}_T 
&-& \frac{w_{qg}^{\rm Asym}}{2}
\Rightarrow 
{\cal O}(Q_\perp^0)
\nonumber \\
w^{qg}_L 
&-& \frac{Q_\perp^2}{Q^2}\,
  \frac{w_{qg}^{\rm Asym}}{2}\
\Rightarrow \,
e_q^2\,\frac{8\pi^2\alpha_s}{3x_1x_2}\, \delta(1-z_1)
\nonumber \\
&&\times
\left[P_{qg}(-z_2)-P_{qg}(z_2)\right]
+{\cal O}(Q_\perp^2)
\nonumber \\
w^{qg}_{\Delta\Delta} 
&-& \frac{1}{2}\,\frac{Q_\perp^2}{Q^2}\,
  \frac{w_{qg}^{\rm Asym}}{2}\
\Rightarrow \,
\frac{1}{2}\, e_q^2\,\frac{8\pi^2\alpha_s}{3x_1x_2}\,\delta(1-z_1)
\nonumber \\
&& \times
\left[P_{qg}(-z_2)-P_{qg}(z_2)\right]
+{\cal O}(Q_\perp^2) .
\label{sf-finite-qg-qt0}
\end{eqnarray}
With $z_1$ and $z_2$ switched, Eq.~(\ref{sf-finite-qg-qt0})  
is also true for the gluon-quark subprocess.
Other than the non-logarithmic finite piece 
(as $Q_{\perp}/Q \rightarrow 0$) in the quark-gluon
contribution to $W_L$ and $W_{\Delta\Delta}$, 
the asymptotic tensor completely removes the leading term 
of the perturbatively calculated helicity structure functions 
as $Q_\perp/Q\to 0$.  

We note that the Lam-Tung relation, $W_L = 2 W_{\Delta \Delta}$,   
is obeyed by the singular terms of the perturbatively calculated helicity 
structure functions.

{\it Polarized initial partons\ }
To gain insight into the uncanceled finite terms in 
$W_L$ and $W_{\Delta\Delta}$ for the $qg$ subprocess, 
we calculate the parton-level 
asymptotic and perturbatively finite helicity structure functions
for ``polarized'' incoming parton states, defined as the states 
with incoming parton polarization projected onto the {\it difference} 
of the parton helicity states~\cite{bqr-long}.  For the quark-antiquark 
subprocess, we find that 
$\Delta w_{q\bar{q}}^{\rm Asym} = w_{q\bar{q}}^{\rm Asym}$, 
whereas for the quark-gluon subprocess  
\begin{equation}
\frac{\Delta w_{qg}^ {\rm Asym}}{2}
\approx 
e_q^2\,\frac{8\pi^2\alpha_s}{3x_1x_2}\,\frac{Q^2}{Q_\perp^2}\,
\Delta P_{qg}(z_2)\delta(1-z_1) ,
\label{}
\end{equation}
with 
\begin{equation}
\Delta P_{qg}(z) 
= T_R 
\left[ z^2 - (1-z)^2 \right]\, .
\label{dpqg}
\end{equation}
We note that $\Delta w_{qg}^{\rm Asym}$ is different from 
$w_{qg}^{\rm Asym}$ in Eq.~(\ref{w-asym-parton}).
The finite contributions in the perturbatively 
computed parton-level helicity structure functions are 
\begin{eqnarray}
\Delta w^{qg}_T 
&-& \frac{\Delta w_{qg}^{\rm Asym}}{2}\
\Rightarrow \,
{\cal O}(Q_\perp^0)
\nonumber \\
\Delta w^{qg}_L 
&-& \frac{Q_\perp^2}{Q^2}\,
  \frac{\Delta w_{qg}^{\rm Asym}}{2}\
\Rightarrow \,
e_q^2\,\frac{8\pi^2\alpha_s}{3x_1x_2}\, \delta(1-z_1)
\nonumber \\
&& \times
\left[\Delta P_{qg}(-z_2)-\Delta P_{qg}(z_2)\right]
+{\cal O}(Q_\perp^2)
\nonumber \\
\Delta w^{qg}_{\Delta\Delta} 
&-& \frac{1}{2}\,\frac{Q_\perp^2}{Q^2}\,
    \frac{\Delta w_{qg}^{\rm Asym}}{2}\
\Rightarrow \,
\frac{1}{2}\, e_q^2\,\frac{8\pi^2\alpha_s}{3x_1x_2}\,\delta(1-z_1) 
\nonumber \\
&& \times
\left[\Delta P_{qg}(-z_2)-\Delta P_{qg}(z_2)\right]
+{\cal O}(Q_\perp^2) .
\label{sf-finite-dqg-qt0}
\end{eqnarray}

The uncanceled finite term in the helicity structure functions 
$W_L$ and $W_{\Delta\Delta}$ is proportional to
\begin{equation}
P_{qg}(-z_2)-P_{qg}(z_2) = 4 z_2 T_R\, , 
\end{equation}
for unpolarized initial partonic states, and to  
\begin{equation}
\Delta P_{qg}(-z_2)-\Delta P_{qg}(z_2) = - 4 z_2 T_R\, ,
\end{equation}
for the ``polarized'' initial partonic states.
Therefore, the uncanceled finite term in $W_L$ and $W_{\Delta\Delta}$ in 
the limit $Q_\perp/Q\to 0$ can also be removed by the asymptotic term 
if the incoming quark and gluon have the same helicity (both positive 
or negative).  We also observe that, at this order, 
the uncanceled term in the quark-gluon subprocess 
is proportional to the helicity flipping splitting function, 
\begin{equation}
P_{q^-g^+}(z)=P_{q^+g^-}(z) = T_R (1-z)^2\, .
\end{equation}
The finite term as $Q_{\perp}/Q \rightarrow 0$ 
for the quark-antiquark subprocess at 
this order is removed completely by the asymptotic term 
since the helicity flipping 
splitting function for the quark vanishes at this order, 
$P_{q^-q^+}(z)=P_{q^+q^-}(z) = 0$.

Our second result in this Letter is that transverse momentum
dependent factorization for the full hadronic tensor, which is the
basis for the Collins-Soper-Sterman $b$-space resummation, 
holds at leading power in the $Q_\perp/Q$ 
expansion.  Breaking of factorization occurs at subleading power but 
only in the helicity flipping channel and does not supply 
leading logarithmic terms.

{\bf Resummation\ }
As shown above, as $Q_\perp/Q\to 0$, all leading large logarithmic 
terms in the helicity structure functions 
$W_T$, $W_L$, and $W_{\Delta\Delta}$ are included 
in one asymptotic function, $W^{\rm Asym}$, the same asymptotic 
function as in the resummed angular-integrated  
Drell-Yan cross section.  
Resummation of the large logarithmic terms of the Drell-Yan 
helicity structure functions can therefore be expressed in terms  
of the resummed contribution to the angular-integrated 
Drell-Yan cross section.
Referring to Eq.~(\ref{st-asym-cs}), we obtain the resummed
contribution to the helicity structure functions in the Collins-Soper
frame as 
\begin{eqnarray}
W_T^{\rm Resum}
&=& 
\left(1-\frac{1}{2} \frac{Q_\perp^2/Q^2}{1+Q_\perp^2/Q^2} \right)
\frac{W^{\rm Resum}}{2} \, ,
\nonumber \\
W_L^{\rm Resum}
&=& 
\frac{Q_\perp^2/Q^2}{1+Q_\perp^2/Q^2}\,
\frac{W^{\rm Resum}}{2}\, ,
\nonumber \\
W_{\Delta\Delta}^{\rm Resum}
&=& 
\frac{1}{2}\,\frac{Q_\perp^2/Q^2}{1+Q_\perp^2/Q^2}\,
\frac{W^{\rm Resum}}{2} \, .
\label{st-resum-cs}
\end{eqnarray}
All depend on the same QCD resummed expression $W^{\rm Resum}$ that  
pertains to the angular-integrated Drell-Yan cross 
section~\cite{css-resum,bqr-long}.
We remark that resummation deals with the 
logarithmically divergent terms in the perturbatively 
calculated structure functions.  
There are remaining finite terms as defined in Eq.~(\ref{w-finite}) 
that make up the full contribution to the structure functions.

Equation~(\ref{st-resum-cs}) shows that the Lam-Tung relation, 
$W_L=2W_{\Delta\Delta}$, is satisfied by the resummed contribution to
all orders in $\alpha_s$. This result is a direct consequence of 
current conservation that led us to the asymptotic tensor in 
Eq.~(\ref{w-asym-cs}). Furthermore, since the resummed contribution 
dominates the cross section in the region of small and intermediate 
values of $Q_\perp$, we expect that violation of the Lam-Tung relation  
as a function of $Q_\perp$ should be relatively small.  This third
result of our investigation is consistent with what is found in
perturbative calculations at order $\alpha_s^2$~\cite{nlo}, but it
is shown here to be true to all orders in $\alpha_s$.  
Data also show reasonable agreement with the Lam-Tung relation for
moderate values of $Q_\perp$ \cite{data}.

{\bf Summary and Discussion\ }  
In this Letter, we investigate the transverse momentum 
$Q_\perp$ dependence of the angular distribution of
leptons from the decay of a massive virtual photon produced in hadron
collisions.  We work in terms of the four independent helicity
structure functions that characterize this angular distribution.  We
show for the first time that the $Q_{\perp}$
dependence of the helicity structure functions $W_T$, $W_L$, and
$W_{\Delta \Delta}$ is specified by the same all-orders resummed
$Q_\perp$ distribution that characterizes the angular
integrated cross section, up to $Q_{\perp}$ dependent kinematic
factors that we list explicitly in Eq.(\ref{st-resum-cs}).  One
consequence of this work is the demonstration that the Lam-Tung
relation, $W_L = 2 W_{\Delta \Delta}$, between the longitudinal and
the double-spin-flip structure functions is obeyed by the resummed
cross sections as a function of $Q_{\perp}$ and that this relation is
a direct result of the requirement of current conservation. 

In this Letter, we work entirely in the collinear factorization
approach and resum large logarithmic perturbative contributions 
when $Q_\perp/Q\to 0$.  When $Q_\perp$ is very small, one might study
the angular distribution in terms of transverse momentum dependent 
parton distributions \cite{bdw-dy}.

In further work, we intend to examine the $Q_{\perp}$ dependence of
$W$ and $Z$ boson production, where parity violating terms introduce
additional helicity structure functions.

{\bf Acknowledgments\ }
E.L.B. is supported by the U.~S.~Department of Energy, Division of High
Energy Physics, under Contract No.\ DE-AC-02-06CH11357.  
J-W.Q. is supported in part by the U.~S.~Department of Energy under
Grant No.\ DE-FG02-87ER40371 and in part by the Argonne University of
Chicago Joint Theory Institute (JTI) Grant
03921-07-137.  R.A.R. is supported in part by the U.~S.~Department of 
Energy under Grant No.\ DE-FG02-87ER40371.
We are grateful to John T. Donohue, Bordeaux, for
valuable communications.


\end{document}